\documentclass[12pt]{spieman}  
\usepackage{amsmath,amsfonts,amssymb}
\usepackage{graphicx}
\usepackage{setspace}
\usepackage{tocloft}
\usepackage{lineno}
\title{Nonlinear virtual lens for programmable and multispectral infrared upconversion imaging}

\author[a]{Ze Zheng}
\author[b]{Olga Sergaeva}
\author[b,c]{Davide Rocco}
\author[d]{Yuchen Zhang}
\author[d]{Guoquan Zhang}
\author[a,*]{Mohsen Rahmani}
\author[b,c,*]{Costantino De Angelis}
\author[a,*]{Lei Xu}
\affil[a]{Advanced Optics and Photonics Laboratory, Department of Engineering, School of Science \& Technology, Nottingham Trent University, Nottingham NG11 8NS, UK}
\affil[b]{Department of Information Engineering, University of Brescia, via Branze 38, 25123 Brescia, Italy}
\affil[c]{National Institute of Optics, Consiglio Nazionale delle Ricerche, via Branze 45, 25123 Brescia, Italy}
\affil[d]{The MOE Key Laboratory of Weak Light Nonlinear Photonics, TEDA Applied Physics Institute and School of Physics, Nankai University, Tianjin 300457, China}

\cftpagenumbersoff{figure}
\cftpagenumbersoff{table} 
\begin{document} 
\maketitle

\begin{abstract}
Conventional infrared (IR) imaging techniques depend on IR cameras based on narrow-bandgap semiconductors, which offer limited spectral bandwidth, coupled with a separate lens.
Recently, advances in nonlinear flat optics have opened a novel pathway for converting IR signals into the visible through nonlinear generations, enabling the direct visualisation of IR images using standard visible cameras. However, the narrow spectral bandwidth and the requirement for an additional lens remain the key challenges. Here, we address both issues via a novel adaptive and multifunctional IR-to-visible imaging platform offering tunable bandwidth and focusing simultaneously. We utilise sum-frequency generation (SFG) to convert IR light into the visible, by introducing a pump beam modulated by a spatial light modulator (SLM) to construct a virtual metalens enabling precisely controlled focusing of the generated nonlinear optical field. As a result, we demonstrate both theoretically and experimentally an optical focusing mechanism with a tunable focal length, achieved by varying the pump and signal wavelengths and modulating the phase distribution. Furthermore, since the focal length depends on the input signal wavelength, the imaging plane position varies accordingly, indicating a promising potential for the multispectral IR imaging applications. Our upconversion platform delivers SLM-controlled, programmable multispectral focusing for next-generation IR imaging, opening new avenues in the fields of computational and multispectral imaging techniques.
\end{abstract}


{\noindent \footnotesize\textbf{*}Mohsen Rahmani,  \linkable{mohsen.rahmani@ntu.ac.uk} }

{\noindent \footnotesize\textbf{*}Costantino De Angelis,  \linkable{costantino.deangelis@unibs.it} }

{\noindent \footnotesize\textbf{*}Lei Xu,  \linkable{lei.xu@ntu.ac.uk} }

\begin{spacing}{2}   

\section{Introduction}
\label{sect:intro}  
IR imaging has become a cornerstone technology for modern industry, enabling applications in medicine\cite{lahiri2012medical}, sensing\cite{norton1991infrared}, night vision\cite{waxman1997color,gade2014thermal}, and building inspection\cite{kylili2014infrared}. The conventional detection techniques of IR light rely on the use of IR cameras. However, IR cameras typically provide lower spatial resolution than visible sensors\cite{bianconi2020recent}, lack colour information\cite{gade2014thermal}, and increase the system cost\cite{bianconi2020recent}, particularly as cryogenic cooling is required for many IR cameras to improve their sensitivity\cite{rogalski2023infrared}. Moreover, traditional IR cameras rely on narrow-bandgap semiconductors, meaning that a given IR camera is sensitive only to a specific spectral range and cannot cover the entire IR range, as efficient IR photon detection is constrained by the semiconductor bandgap \cite{rogalski2003infrared}. These constraints 
motivate the development of nonlinear frequency-conversion approaches that can 
translate IR information into the visible, enabling IR detection with conventional visible cameras 
leveraging the strengths of both systems\cite{camacho2021infrared,huang2022wide,zheng2023third,zheng2024broadband}.

One of the fundamental processes in nonlinear optics, nonlinear frequency conversion, is the physical process by which light at fundamental optical frequencies generates coherent output at the sum or difference frequencies within nonlinear materials, including sum-frequency generation (SFG) and its degenerate case, second-harmonic generation (SHG), as well as higher-order processes such as 
four-wave mixing (FWM)\cite{franken1961generation,boyd2008nonlinear}, etc. A wide range of platforms for such nonlinear optical processes has been extensively studied, including bulk nonlinear crystals\cite{franken1961generation,dmitriev2013handbook}, optical fibres\cite{agrawal2011nonlinear,agrawal2000nonlinear}, and integrated waveguides\cite{lin2007nonlinear,luo2019semi}. Among these, satisfying the phase-matching (momentum-conservation) condition is essential for efficient generation\cite{boyd2008nonlinear,agrawal2000nonlinear,espinola2004raman}, but it also restricts operation to a relatively narrow spectral bandwidth and specific wave-vector (propagation) configurations of the input waves\cite{fejer2002quasi}. These constraints substantially limit the practical use of these conventional platforms.

Flat-optical platforms (including nonlinear metasurfaces) relax the conventional phase-matching constraints for efficient nonlinear generation\cite{vabishchevich2023nonlinear,rahmani2018nonlinear,gigli2022all}. Unlike bulk crystals, fibres, or waveguides, where one balances dispersion over millimetre- to centimetre-scale interaction lengths, the physical thickness in flat optics is typically subwavelength\cite{vabishchevich2023nonlinear}. Consequently, the accumulated phase-mismatch term $\Delta kd$ (with $\Delta k$ the phase mismatch and $d$ the propagation distance) remains $\ll 1$ and can often be neglected\cite{krasnok2018nonlinear,almeida2016subwavelength}. This ultrathin platform supports broadband, angle-tolerant nonlinear generation\cite{krasnok2018nonlinear,almeida2016subwavelength}, which has broadened the application space of flat optics, including optical data processing\cite{pors2015analog,zhou2020flat,de2023analog,gao2021recent}, optical computing\cite{de2023analog,cotrufo2025nonlinear}, and infrared imaging\cite{camacho2021infrared,huang2022wide,zheng2023third,zheng2024broadband}.

Recently, the IR imaging technique based on nonlinear generation has been demonstrated with visible cameras in combination with nonlinear metasurfaces\cite{camacho2021infrared,zheng2023third,zheng2024broadband} and nanowire arrays\cite{fedorov2020}. In these systems, IR scenes are efficiently up-converted to the visible, enabling conventional visible cameras to perform IR detection. Notably, broadband IR imaging via four-wave mixing has been achieved, highlighting the advantages of flat-optical platforms for imaging\cite{zheng2024broadband}. 
On the other hand, there is a growing need to adapt in real time to scene dynamics and system fluctuations, enabling rapid aberration correction\cite{booth2014adaptive,zhang2023adaptive,hong2025long}, dynamic focusing\cite{nikolenko2008slm,hernandez2016three}, and agile beam steering\cite{karabey2013electronic,goltsos1990agile}. Real-time reconfigurable wavefront control has become both essential and challenging. Spatial light modulators (SLMs), which dynamically manipulate phase and amplitude in a two-dimensional pixel array, have emerged as powerful tools for agile wavefront engineering\cite{nikolenko2008slm,yang2023review}. SLM-driven optical systems are now widely used in various domains, including computational imaging\cite{rosen2024roadmap,barbastathis2019use}, focusing through random scattering media\cite{hong2024robust}, and physical computing\cite{fu2024optical,khonina2024exploring}.

\begin{figure}
\begin{center}
\begin{tabular}{c}
\includegraphics[width=\linewidth]{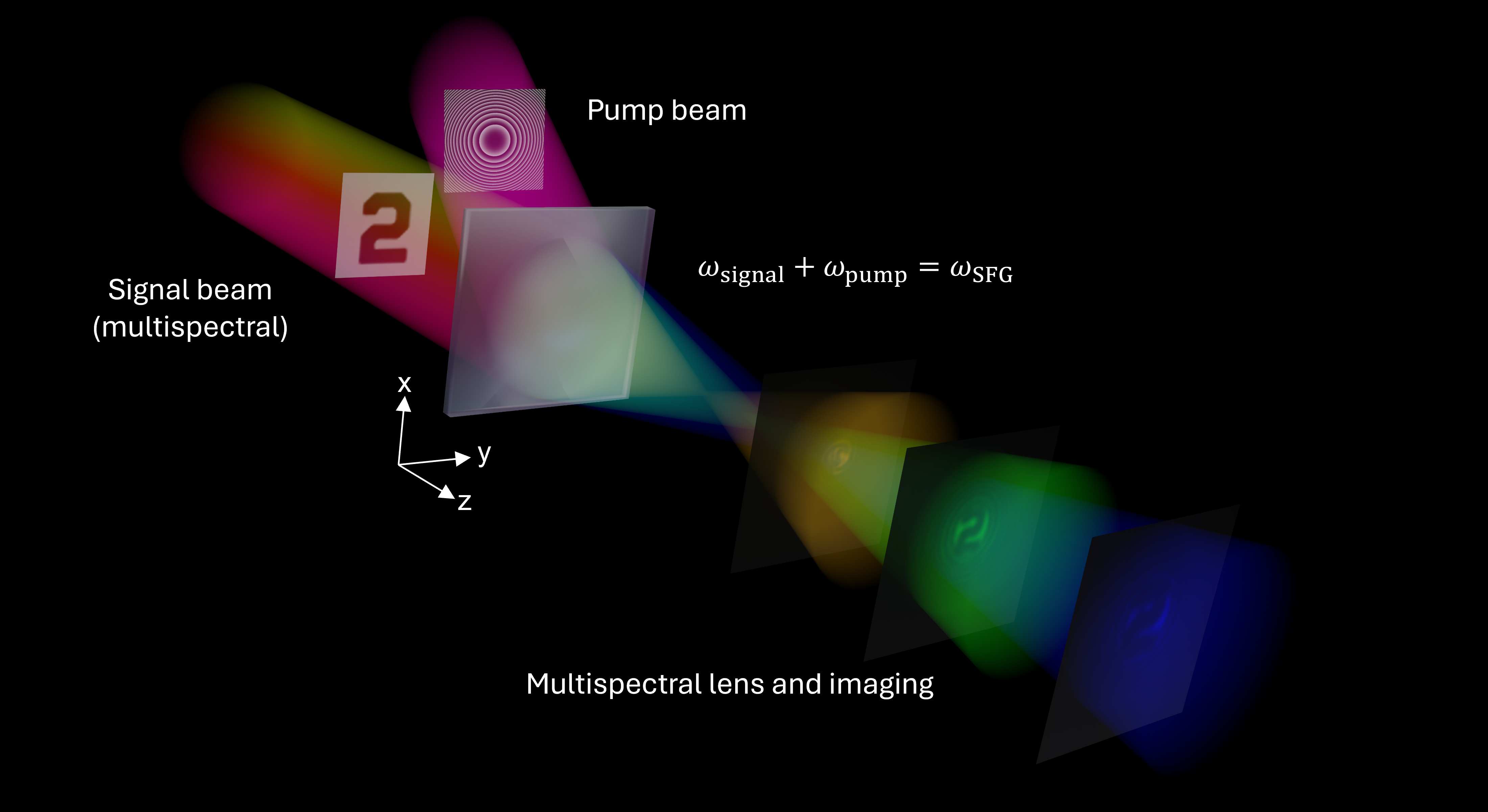}
\end{tabular}
\end{center}
\caption 
{ \label{fig:dia}
The schematic illustrates our dynamic IR focusing and imaging technique based on SFG with a visible camera and an SLM. The number ‘2’ in the signal arm marks the imaging target. In the pump arm, the FZL phase, modulated via the SLM, provides the required phase distribution. The yellow, green, and purple light cones denote SFG emission at different wavelengths corresponding to different pump wavelengths, which focus at distinct focal lengths. Consequently, image planes corresponding to different wavelengths are formed at different axial positions. }
\end{figure} 

In this work, we propose a virtual lens platform based on SFG process that combines IR lensing and imaging together, achieving real-time programmable lensing and multispectral IR imaging via nonlinear generation in a thin lithium niobate film. Using an SLM to encode the real-time phase profile onto a pump beam, we demonstrate dynamically tunable IR focusing and imaging that can be directly captured by visible cameras, as shown in the schematic diagram (Fig.~\ref{fig:dia}). A 600-nm-thick Lithium niobate (LN, LiNbO$_3$) film serves as the nonlinear platform converting IR signals to the visible. A collimated Gaussian signal beam carries imaging information. We imprint a Fresnel zone lens (FZL) phase onto the pump beam with the SLM, which introduces the SFG interaction and induces nonlinear focusing of the SFG emission. Consequently, a signal beam at near-infrared region that carries imaging information is up-converted and recorded by a standard visible camera.   
As the effective focal length depends on the wavelengths of both the signal and the pump, the imaging plane shifts when tuning these wavelengths, thereby enabling multispectral imaging within the same architecture. Our approach opens a new avenue for programmable, multispectral infrared imaging based on nonlinear generation, advancing the applications of nonlinear imaging and wavefront shaping.

\section{Results}
We begin by introducing the concept of nonlinear lensing with an idealised theoretical model. First, an ultrathin lithium niobate film is utilised as the second-order nonlinear medium. For simplicity, we ignore the light longitudinal propagation within the film and phase-matching effects due to the negligible thickness of our film.
The generated SFG and SHG intensities are estimated by treating the LN film as a sheet of second-order polarisation, as described by the equation\cite{boyd2008nonlinear}: 
\begin{equation}
\label{eq:fov}
P_i^{(2)}(\omega_1 + \omega_2) = 2\,\varepsilon_0 \sum_{j,k} d_{ijk}\, E_j{(\omega_1)} E_k{(\omega_2)} \, ,
\end{equation}
where $\varepsilon_0$ is the vacuum permittivity, $P_i^{(2)}$ is the $i$-th Cartesian component of the second-order nonlinear polarisation, 
$E_j{(\omega_1)}$ and $E_k{(\omega_2)}$ are the Cartesian components of the incident electric fields (for SFG: $\omega_1 = \omega_\mathrm{pump}$, $\omega_2 = \omega_\mathrm{signal}$; for SHG: $\omega_1 = \omega_2 = \omega_\mathrm{pump}$) 
and $d_{ijk}$ are the elements of the second-order nonlinear tensor. The entire system is defined in a Cartesian coordinate system: the $x$- and $y$-directions lie in the film plane, and the z-direction is normal to the lithium niobate film.

We consider a 1550-nm pump beam whose optical axis is normal to the lithium niobate film. A Fresnel zone lens (FZL) phase distribution is encoded at the SLM plane and projected onto the film plane through a 4f system, and is defined by~\cite{wood1905physical}:
\begin{equation}
  \phi_{\mathrm{cont}}(x,y)
  = \operatorname{mod}\!\Bigl(\tfrac{2\pi}{r_1^2}\,(x^2+y^2),\,2\pi\Bigr)
  = 2\pi\,\mathrm{frac}\!\Bigl(\tfrac{x^2+y^2}{r_1^2}\Bigr),
  \label{eq:cont-phase}
\end{equation}
where \(\mathrm{frac}(x)=x-\lfloor x\rfloor\) denotes the fractional part and \(\operatorname{mod}(a,2\pi)\) wraps the phase into \([0,2\pi)\), and \(r_1\) is the radius of the first half-wave FZL zone. The phase distribution at the film plane enables the focusing of the pump beam with the focal length, which can be calculated as~\cite{wood1905physical}:
\begin{equation}
  f_{\mathrm{FZL}} = \frac{r_1^2}{2\lambda_\mathrm{pump}}
  =\frac{(n_{\min} p f_2 / f_1)^2}{2\lambda_\mathrm{pump}}.
  \label{eq:f-from-r1}
\end{equation}

\begin{figure}
\begin{center}
\begin{tabular}{c}
\includegraphics[width=\linewidth]{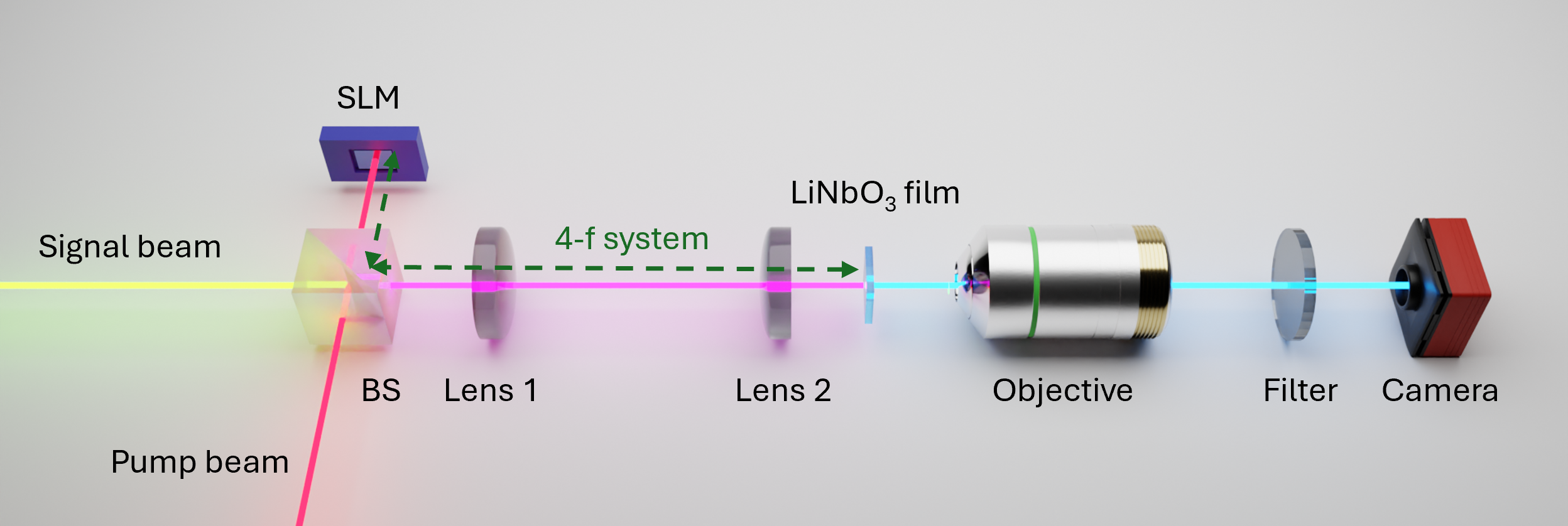}
\end{tabular}
\end{center}
\caption 
{ \label{fig:setup}
The schematic shows the optical setup used to achieve and measure the IR focusing and imaging. (BS is the 50:50 beam splitter.)} 
\end{figure} 

A collimated Gaussian signal beam is also introduced with normal incidence on the lithium niobate film. LiNbO$_3$, as a second-order nonlinear medium, with the pump and signal incidences, supports three nonlinear optical processes arising from the pump and signal interactions: two SHG ($\omega_\mathrm{SHGs}=2\omega_\mathrm{signal}$ and $\omega_\mathrm{SHGp}=2\omega_\mathrm{pump}$) and one SFG ($\omega_\mathrm{SFG}=\omega_\mathrm{signal}+\omega_\mathrm{pump}$). Among these, nonlinear focusing occurs in processes involving the pump beam, since the FZL phase is transferred by the pump field. The SHG process $\omega_\mathrm{SHGs}=2\omega_\mathrm{signal}$ will not be discussed further, as it does not exhibit focusing behaviour. In this work, we primarily investigate the SHG induced by the pump beam ($\omega_\mathrm{SHGp}=2\omega_\mathrm{pump}$) and the SFG induced by both pump and signal beams ($\omega_\mathrm{SFG}=\omega_\mathrm{signal}+\omega_\mathrm{pump}$).
Unless stated otherwise, throughout the discussion, the term “SHG” used below specifically refers to the pump-induced second-harmonic process ($\omega_\mathrm{SHGp}=2\omega_\mathrm{pump}$).

The optical setup used to realise the theoretical model above is illustrated in Fig. \ref{fig:setup}. Two femtosecond lasers provide the signal and pump beams. The signal beam passes sequentially through a beam splitter (BS), lens 1, lens 2, and the lithium niobate film, where, in the presence of the pump, it contributes to SFG. The pump beam enters the BS along a different path, is directed to and reflected by the SLM, and then returns to the BS, where it is combined with the signal. After beam combination, the pump follows the same optical path (from lens 1, lens 2 to the lithium niobate film) and, within the film, interacts to generate the nonlinear signals. After spectrally filtering out the pump and signal beams, the generated nonlinear emission is collected by a 20$\times$ objective and detected by a CMOS camera. The SLM modulates the pump beam by imposing an FZL phase distribution, which is projected to the lithium niobate film through a 4f imaging system formed by the SLM, lens 1, and lens 2.
The objective is mounted on a computer-controlled electric motion stage. The 3D intensity distributions of optical field is measured via moving the motion stage as well as the objective. 

The relationship between the induced nonlinear polarisation of SFG and the signal and pump electric fields can be simplified as $\mathbf{P}^{\mathrm{SFG}}(\omega_\mathrm{signal}+\omega_\mathrm{pump}) \propto \boldsymbol{\chi}^{(2)} :
\big[ \mathbf{E}(\omega_\mathrm{signal})\mathbf{E}(\omega_\mathrm{pump}) \big]$, where $\boldsymbol{\chi}^{(2)}$ is the second-order nonlinear susceptibility tensor of lithium niobate, and “:” denotes a double contraction over the field indices. Based on SFG, the phase and amplitude distributions of the pump and signal can be transferred via this relationship, thus determining the nonlinear wavefront. Here, we modulate the pump beam with a FZL phase distribution, allowing it to be transferred to the nonlinear emissions and thus enabling its nonlinear focusing. Based on Eq.~\ref{eq:f-from-r1}, the focal length of the SFG focusing with the projection of the 4-f system can be calculated using
\begin{equation}
  f_{\mathrm{FZL,SFG}} = \frac{(n_{\min} p f_2 / f_1)^2}{2\lambda_\mathrm{SFG}},
  \label{eq:f-from-r1-SFG}
\end{equation}
where \(n_{\min}\) is the pixel number of the first half-wave FZL zone, \(p = 4.5~\mathrm{\mu m}\) is the pixel pitch of the SLM, and \(f_1=150\)~mm and \(f_2=30\)~mm are the focal lengths of lens 1 and lens 2 in the 4-f system, respectively.

While the SHG focusing slightly varies from the SFG focusing, due to the quadratic relationship between the SHG polarisation and the pump electric field, which can be expressed as the equation $\mathbf{P}^{\mathrm{SHG}}(2\omega_\mathrm{pump}) \propto \boldsymbol{\chi}^{(2)} :
\big[ \mathbf{E}(\omega_\mathrm{pump})\mathbf{E}(\omega_\mathrm{pump}) \big]$. The quadratic relationship results in the FZL phase distribution being transferred to the SHG signal two times, and therefore the corresponding focal length must be reduced by a factor of two:
\begin{equation}
  f_{\mathrm{FZL,SHG}} = \frac{1}{2} \frac{(n_{\min} p f_2 / f_1)^2}{2\lambda_\mathrm{SHG}}.
  \label{eq:f-from-r1-SHG}
\end{equation}

We then simulate the nonlinear focusing via modelling the nonlinear electric field using the angular spectrum method (Fourier-domain free-space propagation). The nonlinear polarisation is treated as an effective zero-thickness sheet at the film coordinate, determined by the lithium niobate second-order nonlinear tensor.
The detailed information regarding the simulation can be found in the Supporting Information section I. 

\begin{figure}
\includegraphics[width=1.0\linewidth]{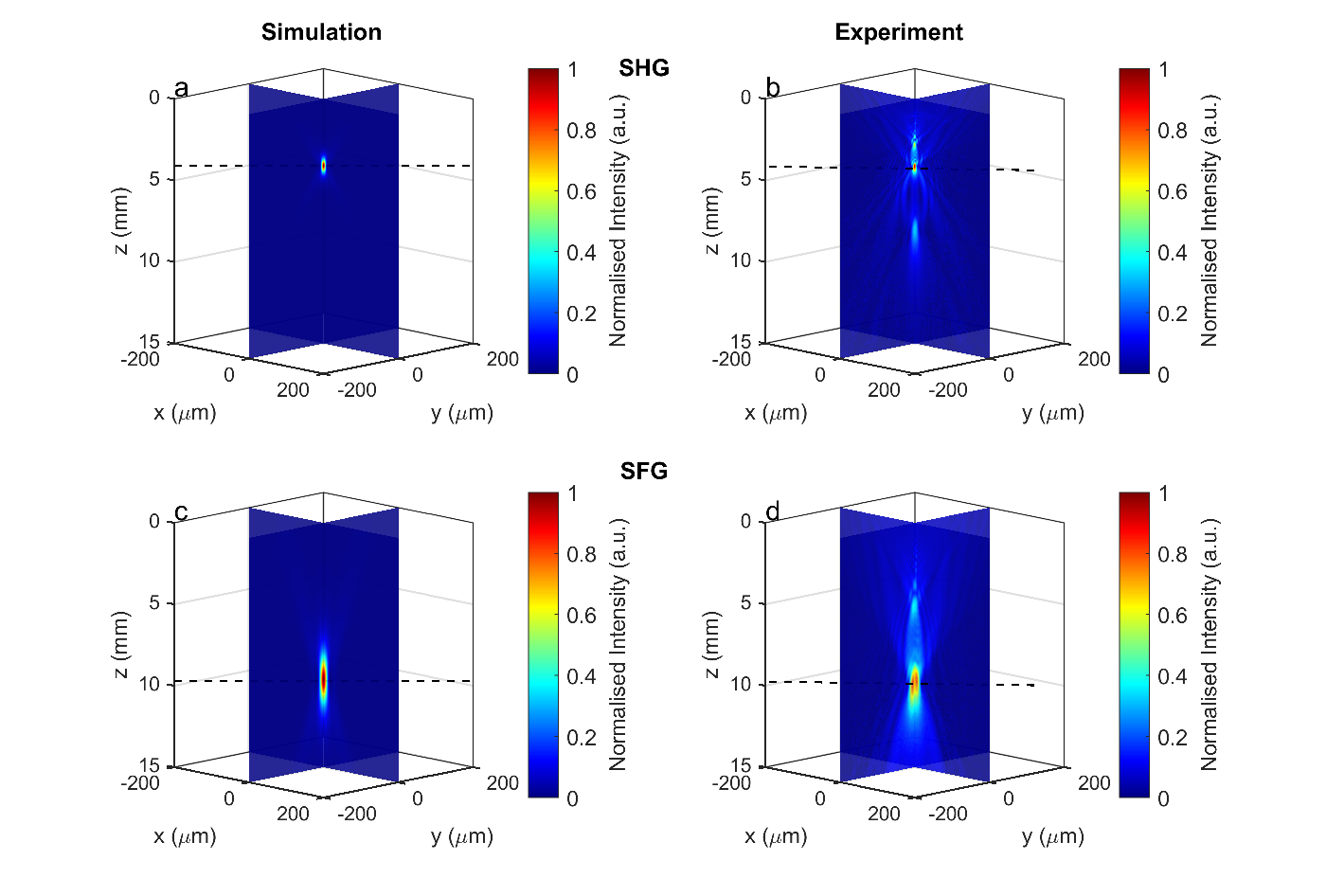}
\caption 
{ \label{fig:simexp}
The simulated (a,c) and measured (b,d) 3D optical field of the SHG (a,b) and SFG (c,d) focusing with the signal wavelength at 1100 nm and the pump wavelength at 1550 nm. The first half-wave zone has a radius of 120 pixels on the SLM in the FZL phase pattern. The normalised intensity exhibits the relative distribution within one optical field, but it does not capture absolute intensity differences across differently focused optical fields.} 
\end{figure}

The comparison between the simulated and measured optical fields of SHG and SFG focusing is exhibited in Fig.~\ref{fig:simexp}. Here, a signal beam with a central wavelength of 1100 nm and a pump beam with a central wavelength of 1550 nm are introduced, thus generating the SHG and SFG fields at 775~nm and 643~nm, consequently. The simulated and experimental optical patterns are in good agreement. In the simulation, where the SHG is generated at a single plane and propagated in free space, the quadratic phase transfer results in the SHG and pump fields focusing at the same axial position. 
The focal length of the simulated (Fig.~\ref{fig:simexp}(a)) and measured (Fig.~\ref{fig:simexp}(b)) SHG focusing is 4.1~mm. The focal length of the SFG focusing in Fig.~\ref{fig:simexp} is 9.7~mm for both simulation (c) and experiment (d).
Moreover, the SFG focal length is almost twice that of the SHG focal length. Given that the SHG and SFG wavelengths are relatively close ($\lambda_\mathrm{SHG}=775~\mathrm{nm};\lambda_\mathrm{SFG}=643~\mathrm{nm}$)  compared with the observed difference in focal lengths, this behaviour demonstrates that the nonlinear interaction governs the transfer of the phase distribution. In particular, it confirms that an additional factor of one half must be included when predicting the SHG focal length.

\begin{figure}
\includegraphics[width=\linewidth]{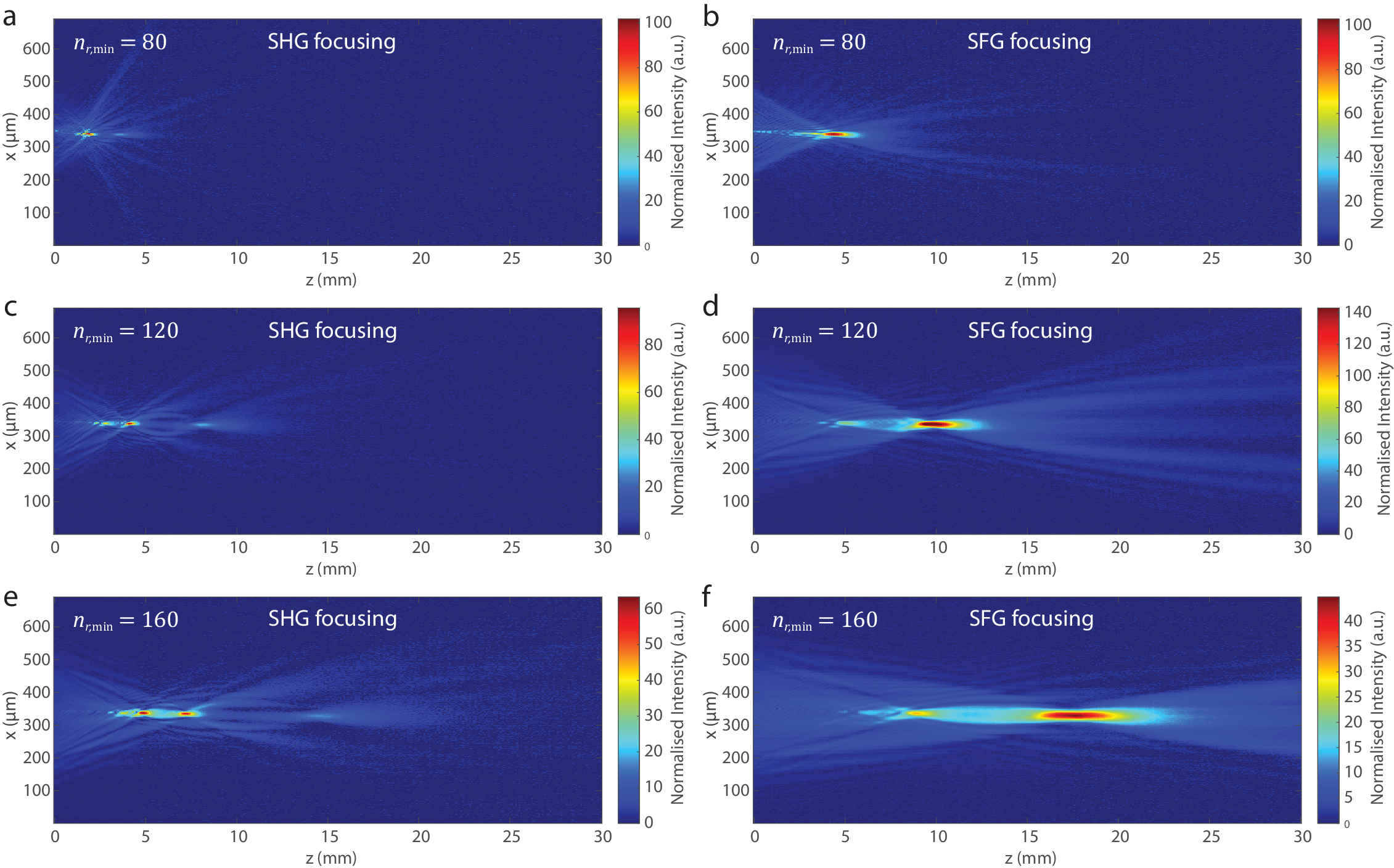}
\caption 
{ \label{fig:cphase}
The measured intensity distributions of optical field in the xz plane of the SHG (a,c,e) and SFG (b,d,f) focusing for first half-wave-zone radii of 80, 120, and 160 pixels on the SLM in the FZL phase pattern: (a, b) 80~px, (c, d) 120~px, (e, f) 160~px. The signal wavelength and pump wavelength are fixed at 1100 and 1550 nm, respectively. 
The intensity was obtained from camera photon counts, corrected for background and normalised by the exposure time, and is therefore shown in relative units. The colour scale is chosen to optimise overall visibility in each subfigure. As a result, the focal region may appear saturated. } 
\end{figure}

According to Eq.~\ref{eq:f-from-r1}, the focal length can be manipulated by tuning the phase distribution and the wavelength. Here, the pixel number refers to the number of SLM pixels defining the radius of the first half-wave Fresnel zone in the FZL phase pattern. 
We first tune the phase distribution by changing the pixel number of the first half-wave FZL zone and then observe the resulting SHG and SFG focusing at different focal lengths (Fig. \ref{fig:cphase} and Supporting Information section II). 
Experimentally, we observe that increasing the pixel number leads to longer focal lengths, for both the SHG (Fig. \ref{fig:cphase}(a,c,e)) and SFG (Fig. \ref{fig:cphase}(b,d,f)), in agreement with the theoretical prediction.

\begin{figure}
\includegraphics[width=\linewidth]{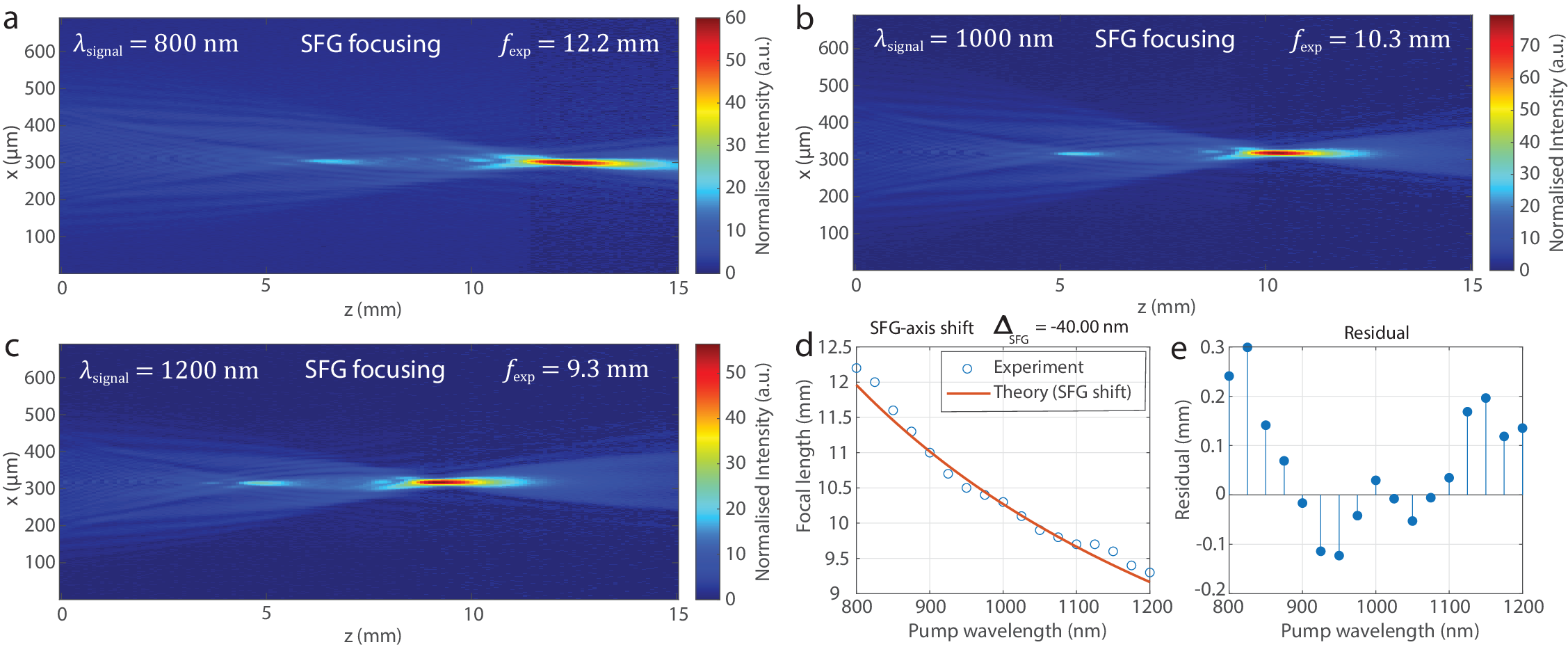}
\caption 
{ \label{fig:cw}
The measured intensity distributions of optical field in the xz plane of the SFG focusing with the signal wavelength at 800 nm (a), 1000 nm (b), 1200 nm (c)~and the pump wavelength at 1550 nm. The measured focal lengths are shown in the top right corner of (a-c). The first half-wave zone has a radius of 120 pixels on the SLM in the FZL phase pattern. The intensity was obtained from camera photon counts, corrected for background and normalised by the exposure time, and is therefore shown in relative units. The colour scale is chosen to optimise overall visibility in each subfigure. As a result, the focal region may appear saturated. (d)~Focal length versus pump wavelength, experiment and theory (Eq.~\ref{eq:f-from-r1-SFG}), with an SFG-axis shift applied to the theory only. Measured focal length (points, mm) is plotted against the pump wavelength and compared to the FZL prediction (line). A constant shift \(\Delta_{\mathrm{sfg}}\) is added only to the theoretical SFG wavelength according to Eq.\ref{eq:f-from-r1-SFG} (\(\lambda_{\mathrm{sfg}}+\Delta_{\mathrm{sfg}}\)). 
(e)~Residuals \(r=f_{\mathrm{exp}}-f_{\mathrm{th}}\) versus pump wavelength for the same fit as (d). 
}
\end{figure}

The focal plane of the SFG emission is determined by both the pump and signal wavelengths. Figures \ref{fig:cw}(a-c) and Supporting Information section III illustrate that the SFG focal length decreases from 12.2 mm to 9.3 mm as the signal wavelength is tuned from 800 nm to 1200 nm. This occurs because each signal wavelength, when combined with a fixed pump wavelength, produces a distinct SFG wavelength and therefore a different focal length. According to Eq.~\ref{eq:f-from-r1-SFG}, the focal length and the wavelength are inversely proportional; consequently, the focal length decreases as the signal wavelength increases. The theoretical focal lengths calculated from Eq.~\ref{eq:f-from-r1-SFG} are compared with the measured values in Fig.~\ref{fig:cw}(d). In the experiment, finite film and substrate thicknesses introduce a small axial shift.
To compensate for offsets arising from the LN film thickness (~600 nm) and the substrate (~1 mm), we apply a constant shift $\Delta_{\mathrm{sfg}}= 40$~nm to the SFG wavelength axis for the theoretical curves only. The measured focal lengths agree closely with the adjusted calculation. The residuals between the theoretical and experimental focal lengths are shown in Fig. \ref{fig:cw}(e).

\begin{figure}
\begin{center}
\begin{tabular}{c}
\includegraphics[width=\linewidth]{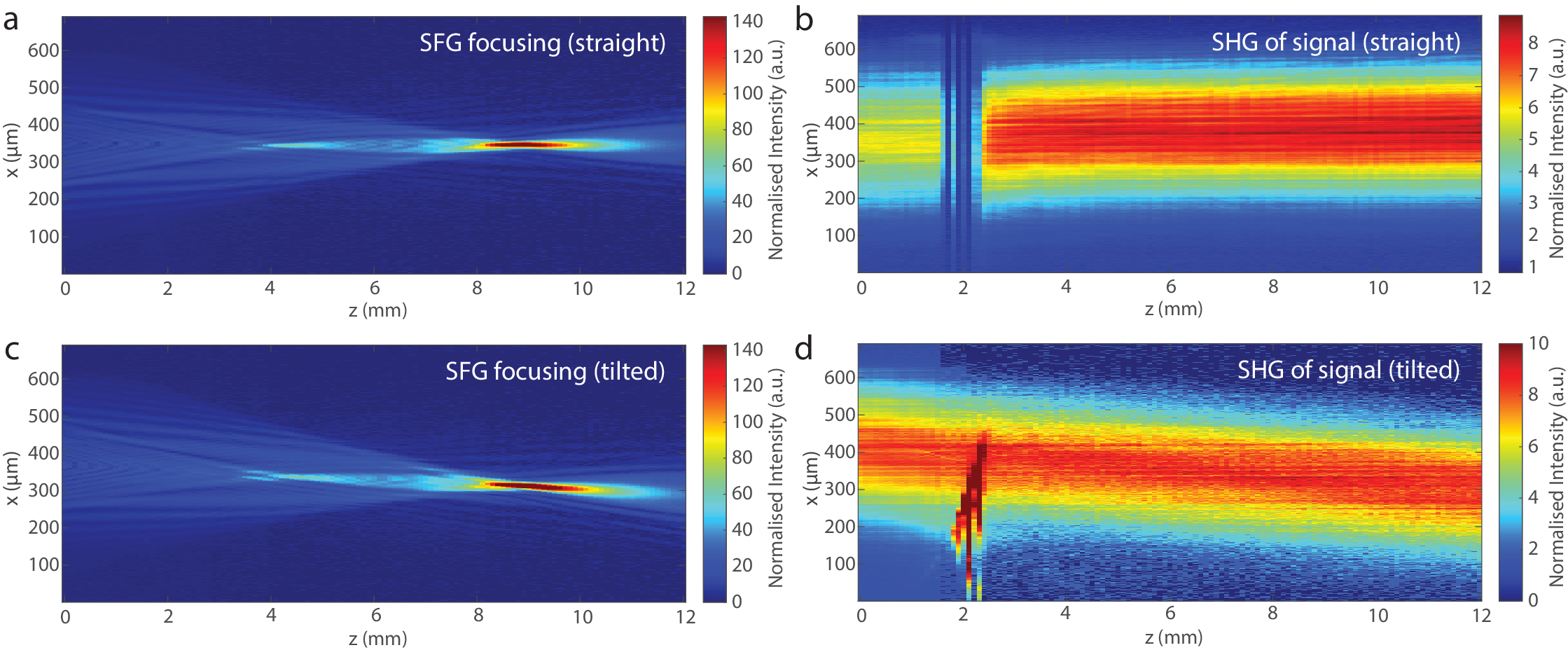}
\end{tabular}
\end{center}
\caption 
{ \label{fig:ck}
The measured intensity distributions of optical field in the xz plane of the SFG focusing and the SHG of the signal incidence under the normal (a,b) and tiled (c,d) incidence of the signal. The first half-wave zone has a radius of 120 pixels on the SLM in the FZL phase pattern. The signal wavelength and pump wavelength are fixed at 1100 and 1550 nm, respectively. The intensity was obtained from camera photon counts, corrected for background and normalised by the exposure time, and is therefore shown in relative units. The colour scale is chosen to optimise overall visibility in each subfigure. As a result, the focal region may appear saturated.} 
\end{figure}

In the preceding text, we prove that the phase distribution is transferred from the fundamental electric field to the nonlinear field based on the nonlinear interaction. Moreover, because phase-matching constraints are relaxed in flat optics, the transverse momentum (wave-vector) is likewise transferred. Figure \ref{fig:ck} reveals this momentum transfer in SFG focusing. We introduce an x-directed tilt on the signal beam and compare it with normal (straight) incidence via observing the SHG of the signal beam (Fig. \ref{fig:ck}(b)(d)). Accordingly, with the pump beam incident normally, the SFG emission is focused on-axis for an untilted signal (Fig. \ref{fig:ck}(a)) and shifts to a corresponding tilt when the signal beam is tilted (Fig. \ref{fig:ck}(c)). This behaviour evidences momentum transfer in the nonlinear process and thus enables IR-to-visible image conversion while preserving most of the incident momentum information.

\begin{figure}
\begin{center}
\begin{tabular}{c}
\includegraphics[width=\linewidth]{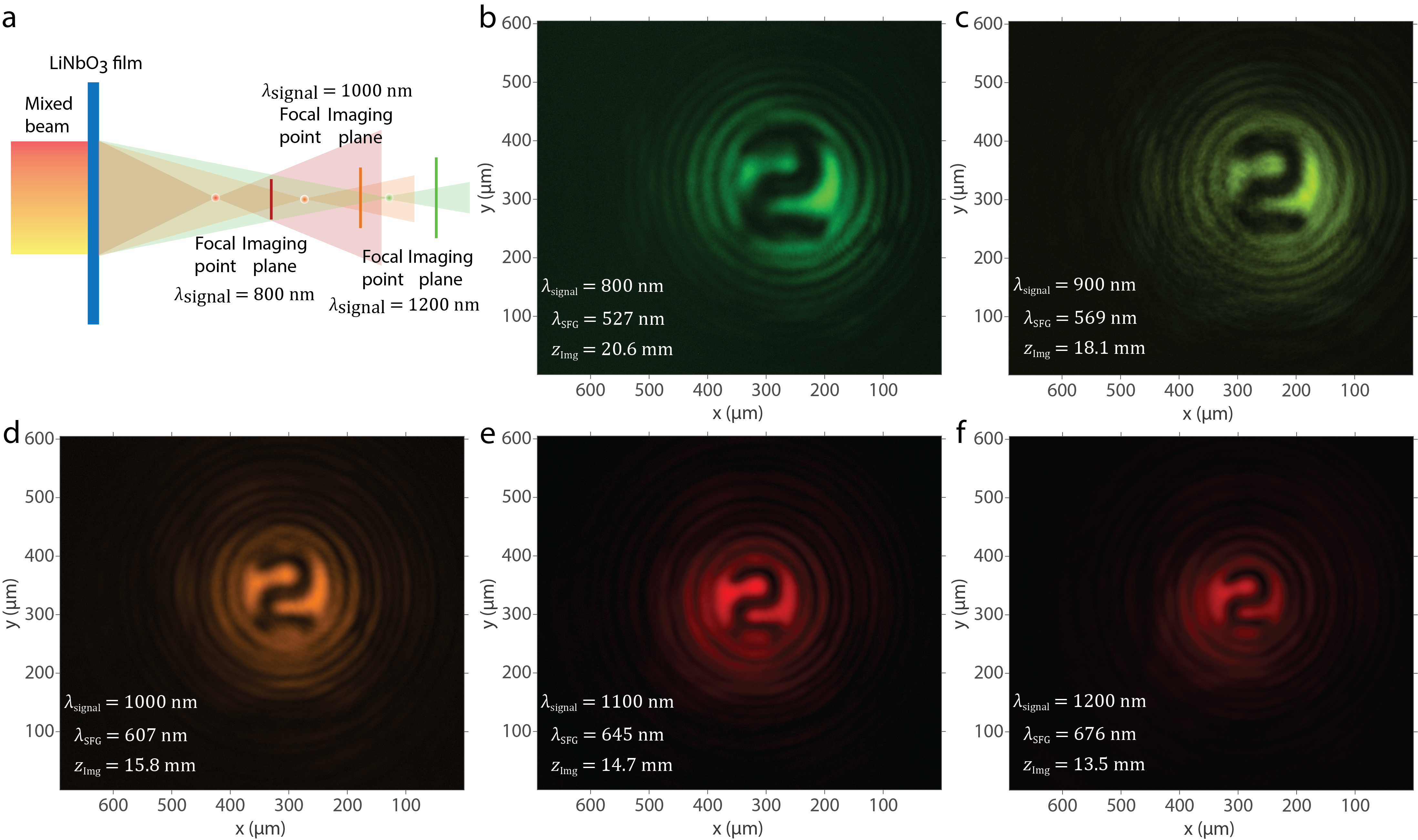}
\end{tabular}
\end{center}
\caption 
{ \label{fig:imaging}
(a)~The schematic demonstrates the multispectral IR imaging.
(b-f)~The converted number '2' images captured by a CMOS scientific camera with different signal wavelengths: (b)~$\lambda_{\mathrm{signal}}=800 \mathrm{nm}$;(c)~$\lambda_{\mathrm{signal}}=900 ~\mathrm{nm}$; (d)~$\lambda_{\mathrm{signal}}=1000 ~\mathrm{nm}$; (e)~$\lambda_{\mathrm{signal}}=1100 ~\mathrm{nm}$; (f)~$\lambda_{\mathrm{signal}}=1200 ~\mathrm{nm}$. 
Here the colours are used to exhibit the SFG radiation spectrum positions. The information on 
the incident signal wavelength, the SFG emission wavelength, and the position of the imaging plane is shown in the bottom left corner of~(b-f).
} 
\end{figure} 

Via utilising the nonlinear focusing of our proposed optical system, the infrared image can be transformed into the visible, allowing for it to be imaged and detected by a visible camera. The input signal at a given wavelength is then transformed via SFG and focused at a specific position according to equation \ref{eq:f-from-r1-SFG}, enabling the corresponding imaging plane to be located at different positions, as illustrated in Fig.~\ref{fig:imaging}(a). An imaging target, number '2', is placed in the signal arm for imaging. The signal wavelength is tuned from 800~nm to 1200~nm. The transformed number '2' images detected by a CMOS scientific camera are illustrated in Fig.~\ref{fig:imaging}(b-f), with colour indicating the SFG wavelength positions. The size of number '2' varies with the position of the imaging plane, due to changes in magnification caused by the tuned focal lengths. Consequently, we achieve the direct conversion of infrared images into visible and locate the imaging planes according to the signal wavelength, demonstrating an initial step towards multispectral nonlinear imaging.

\section{Conclusion}

In summary, we demonstrate dynamic nonlinear focusing enabled by phase transfer in second-order nonlinear interactions. By modulating the pump beam with a Fresnel zone lens (FZL) phase distribution using a spatial light modulator (SLM), the phase is transferred to the generated SHG and SFG fields in a lithium niobate film, resulting in controllable nonlinear focusing. The focal length can be tuned through either the wavelength or the imposed phase distribution. We further show that transverse momentum is transferred from the fundamental field to the nonlinear emission. Building on this mechanism, we demonstrate a broadband infrared imaging approach by placing an imaging target in the signal arm.
Because the focal length varies with the signal wavelength, infrared images at different wavelengths are formed at different imaging planes. These results highlight the role of phase and momentum transfer in nonlinear flat optics and open new opportunities for programmable and multispectral imaging.

\section{Acknowledgment}
Funded by the European Union (ERC, UPIRI, 101170298). Views and opinions expressed are however those of the authors only and do not necessarily reflect those of the European Union or the European Research Council. Neither the European Union nor the granting authority can be held responsible for them.
G.Z. acknowledges the National Key Research and Development Program of China (2022YFA1404604).
Z.Z. and M.R. acknowledge the support from the support from UK Research and Innovation Future Leaders Fellowship (MR/Z000270/1). The authors acknowledge the use of NTU High-Performance Computing cluster Avicenna.

\appendix    


\bibliography{report}   

@book{wood1905physical,
  title={Physical optics},
  author={Wood, Robert Williams},
  year={1905},
  publisher={Macmillan}
}

@article{lahiri2012medical,
  title={Medical applications of infrared thermography: a review},
  author={Lahiri, Barid Baran and Bagavathiappan, Subramaniam and Jayakumar, T and Philip, John},
  journal={Infrared physics \& technology},
  volume={55},
  number={4},
  pages={221--235},
  year={2012},
  publisher={Elsevier}
}

@article{norton1991infrared,
  title={Infrared image sensors},
  author={Norton, Paul R},
  journal={Optical Engineering},
  volume={30},
  number={11},
  pages={1649--1663},
  year={1991},
  publisher={SPIE}
}

@article{waxman1997color,
  title={Color night vision: opponent processing in the fusion of visible and IR imagery},
  author={Waxman, Allen M and Gove, Alan N and Fay, David A and Racamato, Joseph P and Carrick, James E and Seibert, Michael C and Savoye, Eugene D},
  journal={Neural Networks},
  volume={10},
  number={1},
  pages={1--6},
  year={1997},
  publisher={Elsevier}
}

@article{gade2014thermal,
  title={Thermal cameras and applications: a survey},
  author={Gade, Rikke and Moeslund, Thomas B},
  journal={Machine vision and applications},
  volume={25},
  number={1},
  pages={245--262},
  year={2014},
  publisher={Springer}
}

@article{kylili2014infrared,
  title={Infrared thermography (IRT) applications for building diagnostics: A review},
  author={Kylili, Angeliki and Fokaides, Paris A and Christou, Petros and Kalogirou, Soteris A},
  journal={Applied Energy},
  volume={134},
  pages={531--549},
  year={2014},
  publisher={Elsevier}
}

@article{bianconi2020recent,
  title={Recent advances in infrared imagers: toward thermodynamic and quantum limits of photon sensitivity},
  author={Bianconi, Simone and Mohseni, Hooman},
  journal={Reports on Progress in Physics},
  volume={83},
  number={4},
  pages={044101},
  year={2020},
  publisher={IOP Publishing}
}

@article{rogalski2023infrared,
  title={Infrared HOT photodetectors: status and outlook},
  author={Rogalski, Antoni and Kopytko, Ma{\l}gorzata and Hu, Weida and Martyniuk, Piotr},
  journal={Sensors},
  volume={23},
  number={17},
  pages={7564},
  year={2023},
  publisher={MDPI}
}

@article{rogalski2003infrared,
  title={Infrared detectors: status and trends},
  author={Rogalski, Antoni},
  journal={Progress in quantum electronics},
  volume={27},
  number={2-3},
  pages={59--210},
  year={2003},
  publisher={Elsevier}
}

@article{huang2022wide,
  title={Wide-field mid-infrared single-photon upconversion imaging},
  author={Huang, Kun and Fang, Jianan and Yan, Ming and Wu, E and Zeng, Heping},
  journal={Nature communications},
  volume={13},
  number={1},
  pages={1077},
  year={2022},
  publisher={Nature Publishing Group UK London}
}

@article{zheng2024broadband,
  title={Broadband infrared imaging governed by guided-mode resonance in dielectric metasurfaces},
  author={Zheng, Ze and Smirnova, Daria and Sanderson, Gabriel and Cuifeng, Ying and Koutsogeorgis, Demosthenes C and Huang, Lujun and Liu, Zixi and Oulton, Rupert and Yousefi, Arman and Miroshnichenko, Andrey E and others},
  journal={Light: Science \& Applications},
  volume={13},
  number={1},
  pages={249},
  year={2024},
  publisher={Nature Publishing Group UK London}
}

@article{zheng2023third,
  title={Third-harmonic generation and imaging with resonant Si membrane metasurface},
  author={Zheng, Ze and Xu, Lei and Huang, Lujun and Smirnova, Daria and Kamali, Khosro Zangeneh and Yousefi, Arman and Deng, Fu and Camacho-Morales, Rocio and Ying, Cuifeng and Miroshnichenko, Andrey E and others},
  journal={Opto-Electronic Advances},
  volume={6},
  number={8},
  pages={220174--1},
  year={2023}
}

@article{agrawal2011nonlinear,
  title={Nonlinear fiber optics: its history and recent progress},
  author={Agrawal, Govind P.},
  journal={JOSA B},
  volume={28},
  number={12},
  pages={A1--A10},
  year={2011},
  publisher={Optica Publishing Group}
}

@incollection{agrawal2000nonlinear,
  title={Nonlinear fiber optics},
  author={Agrawal, Govind P},
  booktitle={Nonlinear Science at the Dawn of the 21st Century},
  pages={195--211},
  year={2000},
  publisher={Springer}
}

@article{lin2007nonlinear,
  title={Nonlinear optical phenomena in silicon waveguides: modeling and applications},
  author={Lin, Q and Painter, Oskar J and Agrawal, Govind P},
  journal={Optics express},
  volume={15},
  number={25},
  pages={16604--16644},
  year={2007},
  publisher={Optica Publishing Group}
}

@article{luo2019semi,
  title={Semi-nonlinear nanophotonic waveguides for highly efficient second-harmonic generation},
  author={Luo, Rui and He, Yang and Liang, Hanxiao and Li, Mingxiao and Lin, Qiang},
  journal={Laser \& Photonics Reviews},
  volume={13},
  number={3},
  pages={1800288},
  year={2019},
  publisher={Wiley Online Library}
}

@incollection{boyd2008nonlinear,
  title={Nonlinear optics},
  author={Boyd, Robert W and Gaeta, Alexander L and Giese, Enno},
  booktitle={Springer Handbook of Atomic, Molecular, and Optical Physics},
  pages={1097--1110},
  year={2008},
  publisher={Springer}
}

@article{espinola2004raman,
  title={Raman amplification in ultrasmall silicon-on-insulator wire waveguides},
  author={Espinola, Richard L and Dadap, Jerry I and Osgood, Richard M and McNab, Sharee J and Vlasov, Yurii A},
  journal={Optics Express},
  volume={12},
  number={16},
  pages={3713--3718},
  year={2004},
  publisher={Optica Publishing Group}
}

@article{franken1961generation,
  title={Generation of optical harmonics},
  author={Franken, PA and Hill, Alan E and Peters, CW and Weinreich, Gabriel},
  journal={Physical review letters},
  volume={7},
  number={4},
  pages={118},
  year={1961},
  publisher={APS}
}

@book{dmitriev2013handbook,
  title={Handbook of nonlinear optical crystals},
  author={Dmitriev, Valentin G and Gurzadyan, Gagik G and Nikogosyan, David N},
  volume={64},
  year={2013},
  publisher={Springer}
}

@article{fejer2002quasi,
  title={Quasi-phase-matched second harmonic generation: tuning and tolerances},
  author={Fejer, Martin M and Magel, GA and Jundt, Dieter H and Byer, Robert L},
  journal={IEEE Journal of quantum electronics},
  volume={28},
  number={11},
  pages={2631--2654},
  year={2002},
  publisher={IEEE}
}

@article{krasnok2018nonlinear,
  title={Nonlinear metasurfaces: a paradigm shift in nonlinear optics},
  author={Krasnok, Alexander and Tymchenko, Mykhailo and Al{\`u}, Andrea},
  journal={Materials Today},
  volume={21},
  number={1},
  pages={8--21},
  year={2018},
  publisher={Elsevier}
}

@article{vabishchevich2023nonlinear,
  title={Nonlinear photonics with metasurfaces},
  author={Vabishchevich, Polina and Kivshar, Yuri},
  journal={Photonics Research},
  volume={11},
  number={2},
  pages={B50--B64},
  year={2023},
  publisher={Chinese Laser Press and Optica Publishing Group}
}

@article{almeida2016subwavelength,
  title={Subwavelength nonlinear phase control and anomalous phase matching in plasmonic metasurfaces},
  author={Almeida, Euclides and Shalem, Guy and Prior, Yehiam},
  journal={Nature communications},
  volume={7},
  number={1},
  pages={10367},
  year={2016},
  publisher={Nature Publishing Group UK London}
}

@article{de2023analog,
  title={Analog image processing with nonlinear nonlocal flat optics},
  author={de Ceglia, Domenico and Al{\`u}, Andrea and Neshev, Dragomir N and De Angelis, Costantino},
  journal={Optical Materials Express},
  volume={14},
  number={1},
  pages={92--100},
  year={2023},
  publisher={Optica Publishing Group}
}

@article{camacho2021infrared,
  title={Infrared upconversion imaging in nonlinear metasurfaces},
  author={Camacho-Morales, Rocio and Rocco, Davide and Xu, Lei and Gili, Valerio Flavio and Dimitrov, Nikolay and Stoyanov, Lyubomir and Ma, Zhonghua and Komar, Andrei and Lysevych, Mykhaylo and Karouta, Fouad and others},
  journal={Advanced Photonics},
  volume={3},
  number={3},
  pages={036002--036002},
  year={2021},
  publisher={Society of Photo-Optical Instrumentation Engineers}
}

@article{fedorov2020,
author = {Fedorov, Vladimir V. and Bolshakov, Alexey and Sergaeva, Olga and Neplokh, Vladimir and Markina, Daria and Bruyere, Stephanie and Saerens, Gr{\'e}goire and Petrov, Mihail I. and Grange, Rachel and Timofeeva, Maria and Makarov, Sergey V. and Mukhin, Ivan S.},
title = {Gallium Phosphide Nanowires in a Free-Standing, Flexible, and Semitransparent Membrane for Large-Scale Infrared-to-Visible Light Conversion},
journal = {ACS Nano},
volume = {14},
number = {8},
pages = {10624-10632},
year = {2020},
doi = {10.1021/acsnano.0c04872},
URL = {https://doi.org/10.1021/acsnano.0c04872
}
}

@article{pors2015analog,
  title={Analog computing using reflective plasmonic metasurfaces},
  author={Pors, Anders and Nielsen, Michael G and Bozhevolnyi, Sergey I},
  journal={Nano letters},
  volume={15},
  number={1},
  pages={791--797},
  year={2015},
  publisher={ACS Publications}
}

@article{zhou2020flat,
  title={Flat optics for image differentiation},
  author={Zhou, You and Zheng, Hanyu and Kravchenko, Ivan I and Valentine, Jason},
  journal={Nature Photonics},
  volume={14},
  number={5},
  pages={316--323},
  year={2020},
  publisher={Nature Publishing Group UK London}
}

@article{cotrufo2025nonlinear,
  title={Nonlinear analog processing with anisotropic nonlinear films},
  author={Cotrufo, Michele and de Ceglia, Domenico and Jung, Hyunseung and Brener, Igal and Neshev, Dragomir and De Angelis, Costantino and Al{\`u}, Andrea},
  journal={Nanophotonics},
  number={0},
  year={2025},
  publisher={De Gruyter}
}

@article{booth2014adaptive,
  title={Adaptive optical microscopy: the ongoing quest for a perfect image},
  author={Booth, Martin J},
  journal={Light: Science \& Applications},
  volume={3},
  number={4},
  pages={e165--e165},
  year={2014},
  publisher={Nature Publishing Group}
}

@article{zhang2023adaptive,
  title={Adaptive optics for optical microscopy},
  author={Zhang, Qinrong and Hu, Qi and Berlage, Caroline and Kner, Peter and Judkewitz, Benjamin and Booth, Martin and Ji, Na},
  journal={Biomedical Optics Express},
  volume={14},
  number={4},
  pages={1732--1756},
  year={2023},
  publisher={Optica Publishing Group}
}

@article{hernandez2016three,
  title={Three-dimensional spatiotemporal focusing of holographic patterns},
  author={Hernandez, Oscar and Papagiakoumou, Eirini and Tanese, Dimitrii and Fidelin, Kevin and Wyart, Claire and Emiliani, Valentina},
  journal={Nature communications},
  volume={7},
  number={1},
  pages={11928},
  year={2016},
  publisher={Nature Publishing Group UK London}
}

@article{nikolenko2008slm,
  title={SLM microscopy: scanless two-photon imaging and photostimulation using spatial light modulators},
  author={Nikolenko, Volodymyr and Watson, Brendon O and Araya, Roberto and Woodruff, Alan and Peterka, Darcy S and Yuste, Rafael},
  journal={Frontiers in neural circuits},
  volume={2},
  pages={393},
  year={2008},
  publisher={Frontiers}
}

@article{yang2023review,
  title={A review of liquid crystal spatial light modulators: devices and applications},
  author={Yang, Yiqian and Forbes, Andrew and Cao, Liangcai},
  journal={Opto-Electronic Science},
  volume={2},
  number={8},
  pages={230026--1},
  year={2023}
}

@article{hong2025long,
  title={Long-range correlation of transmitted light through transmission eigenchannels in disordered media},
  author={Hong, Peilong and Yi, Mingfang and Zhang, Liwei and Liang, Yi and Zhang, Guoquan},
  journal={Physical Review B},
  volume={111},
  number={2},
  pages={024202},
  year={2025},
  publisher={APS}
}

@article{hong2024robust,
  title={Robust multiple focusing through scattering media via feedback wavefront shaping},
  author={Hong, Peilong and Liang, Yi and Zhang, Guoquan},
  journal={Optics \& Laser Technology},
  volume={176},
  pages={110939},
  year={2024},
  publisher={Elsevier}
}

@book{karabey2013electronic,
  title={Electronic beam steering and polarization agile planar antennas in liquid crystal technology},
  author={Karabey, Onur Hamza},
  year={2013},
  publisher={Springer Science \& Business Media}
}

@article{goltsos1990agile,
  title={Agile beam steering using binary optics microlens arrays},
  author={Goltsos, William C and Holz, Michael},
  journal={Optical Engineering},
  volume={29},
  number={11},
  pages={1392--1397},
  year={1990},
  publisher={SPIE}
}

@article{fu2024optical,
  title={Optical neural networks: progress and challenges},
  author={Fu, Tingzhao and Zhang, Jianfa and Sun, Run and Huang, Yuyao and Xu, Wei and Yang, Sigang and Zhu, Zhihong and Chen, Hongwei},
  journal={Light: Science \& Applications},
  volume={13},
  number={1},
  pages={263},
  year={2024},
  publisher={Nature Publishing Group UK London}
}

@article{khonina2024exploring,
  title={Exploring types of photonic neural networks for imaging and computing—a review},
  author={Khonina, Svetlana N and Kazanskiy, Nikolay L and Skidanov, Roman V and Butt, Muhammad A},
  journal={Nanomaterials},
  volume={14},
  number={8},
  pages={697},
  year={2024},
  publisher={MDPI}
}

@article{rosen2024roadmap,
  title={Roadmap on computational methods in optical imaging and holography},
  author={Rosen, Joseph and Alford, Simon and Allan, Blake and Anand, Vijayakumar and Arnon, Shlomi and Arockiaraj, Francis Gracy and Art, Jonathan and Bai, Bijie and Balasubramaniam, Ganesh M and Birnbaum, Tobias and others},
  journal={Applied Physics B},
  volume={130},
  number={9},
  pages={166},
  year={2024},
  publisher={Springer}
}

@article{barbastathis2019use,
  title={On the use of deep learning for computational imaging},
  author={Barbastathis, George and Ozcan, Aydogan and Situ, Guohai},
  journal={Optica},
  volume={6},
  number={8},
  pages={921--943},
  year={2019},
  publisher={Optical Society of America}
}

@article{rahmani2018nonlinear,
  title={Nonlinear frequency conversion in optical nanoantennas and metasurfaces: materials evolution and fabrication},
  author={Rahmani, Mohsen and Leo, Giuseppe and Brener, Igal and Zayats, Anatoly V and Maier, Stefan A and De Angelis, Costantino and Tan, Hoe and Gili, Valerio Flavio and Karouta, Fouad and Oulton, Rupert and others},
  journal={Opto-Electronic Advances},
  volume={1},
  number={10},
  pages={180021--1},
  year={2018}
}

@article{gao2021recent,
  title={Recent advances in optical dynamic meta-holography},
  author={Gao, Hui and Fan, Xuhao and Xiong, Wei and Hong, Minghui and others},
  journal={Opto-Electronic Advances},
  volume={4},
  number={11},
  pages={210030--1},
  year={2021},
  publisher={Editorial Office of Opto-Electronic Advances}
}

@article{gigli2022all,
  title={All-dielectric $\chi$ (2) metasurfaces: recent progress},
  author={Gigli, Carlo and Leo, Giuseppe},
  journal={Opto-Electronic Advances},
  volume={5},
  number={7},
  pages={210093--1},
  year={2022}
}
\bibliographystyle{spiejour}   



\vspace{1ex}


\end{spacing}
\end{document}